\documentclass[pre,twocolumn]{revtex4}
\usepackage{graphicx,amssymb}

\begin{document}
\title{Nonextensive statistical mechanics, anomalous diffusion and central limit theorems \footnote{Invited paper to appear in {\it Milan Journal of Mathematics} (2005).}}

\author{
Constantino Tsallis \thanks{tsallis@cbpf.br}
}

\address{
Santa Fe Institute,
1399 Hyde Park Road,
Santa Fe, New Mexico 87501,  USA\\
and\\
Centro Brasileiro de Pesquisas Fisicas, Rua Xavier Sigaud 150, 
22290-180 Rio de Janeiro-RJ, Brazil.
}
\date{\today}
\begin{abstract}
We briefly review Boltzmann-Gibbs and nonextensive statistical mechanics as well as their connections with Fokker-Planck equations and with existing central limit theorems. We then provide some hints that might pave the road to the proof of a new central limit theorem, which would play a fundamental role in the foundations and ubiquity of nonextensive statistical mechanics. The basic novelty introduced within this conjectural theorem is the {\it generalization of the hypothesis of independence} of the $N$ random variables being summed. In addition to this, we also advance some nonlinear dynamical (possibly exact) relations which generalize the concepts of Lyapunov exponents, entropy production per unit time, and their interconnection as first proved by Pesin for chaotic systems.
\end{abstract}
\maketitle

\section{Introduction}
\label{section_introduction}
As well known, {\it thermodynamics} is the basic branch of physics which focuses on the generic connections between variables (temperature, pressure, volume, energy, entropy and many others) that play an important role in the description of the {\it macroscopic} world. Boltzmann and Gibbs provided a magnificent connection of thermodynamics with the {\it microscopic} world \cite{Boltzmann1872, Gibbs1902}. This connection, normally referred to as {\it Boltzmann-Gibbs} ($BG$) {\it statistical mechanics} (or simply {\it statistical mechanics} since it was basically the only one to be formulated along more than one century), turns out to be the appropriate one for ubiquitous systems in nature. It is based on the following axiomatic expression for the entropy:
\begin{equation}
S_{BG} \equiv\ -k \sum_{i=1}^W p_i \ln p_i \;,
\end{equation}
with
\begin{equation}
\sum_{i=1}^W p_i=1 \;,
\end{equation}
where $p_i$ is the probability associated with the $i^{th}$ microscopic state of the system, and $k$ is Boltzmann constant. In the particular case of equiprobability, i.e., $p_i=1/W$  $(\forall i)$, Eq. (1) yields the celebrated {\it Boltzmann principle} (as referred to by Einstein himself \cite{Einstein10}):
\begin{equation}
S_{BG}=k \ln W \;.
\end{equation}
From now on, and without loss of generality, we shall take $k$ equal to unity. 

For continuous variables, the $BG$ entropy is written as
\begin{equation}
S_{BG} \equiv\ - \int dx \; p(x) \ln p(x) \;\;\;(x \in{\mathbb R^d}),
\end{equation}  
with
\begin{equation}
\int dx\,p(x)=1 \;.
\end{equation}
If $x$ happens to carry physical units, we write Eq. (4) as follows:
\begin{equation}
S_{BG} \equiv\ - \int dx \; p(x) \ln \Bigl[\Bigl(\prod_{r=1}^d \sigma_r \Bigr) p(x)\Bigr] \;,
\end{equation}  
where $\sigma_r>0$ carries the same units as the $r^{th}$ component of the $d$-dimensional variable $x$. Clearly, when $x$ carries no units (i.e., when $x \in{\mathbb R^d}$), we take $\sigma_r=1\;(\forall r)$. In fact, everytime this is possible (and it {\it is} possible most of the time), we shall adapt the physical units in such a way that $\sigma_r=1 \;(\forall r)$ {\it even when $x$ does have physical units}. Consistently, unless otherwise specified, we shall use Eq. (4) for the general continuous case. If we consider the particular case $p(x)=\sum_{i=1}^W p_i \delta(x-x_i)$ (with $\sum_{i=1}^W p_i=1$, $\{x_i\}$ being some set of values, and $\delta(z)$ being Dirac' s delta distribution), Eq. (4) recovers Eq. (1).

For quantum systems, the $BG$ entropic form  is written as
\begin{equation}
S_{BG} \equiv\ -Tr \rho \ln \rho \;,
\end{equation}
with
\begin{equation}
Tr \rho=1 \;,
\end{equation}
$\rho$ being the density operator or matrix. When the $W \times W$ matrix $\rho$ is diagonalized, it shows the set $\{p_i\}$ in its diagonal. In what follows, depending on the context, we shall use either the discrete form (Eqs. (1) and (2)), or the continuous form (Eqs. (4) and (5)), or the matricial form (Eqs. (7) and (8)).

In spite of its tremendous power and usefulness, the $BG$ concepts and statistical mechanics appear to be {\it not universally applicable}. Indeed, there is a plethora of natural and artificial systems (see, for instance, \cite{GellmannTsallis04} and references therein) for which they do not provide the adequate mathematical frame for handling physically relevant quantities. This fact started being explicitly recognized at least as early as in 1902 by Gibbs himself: see page 35 of \cite{Gibbs1902}, where he addresses anomalies related to systems such as gravitation.
A formalism becomes therefore desirable which would address such anomalous systems. A vast class of them (although surely not {\it all} of them) appears to be adequately discussed within a generalization of the $BG$ theory, frequently referred to as {\it nonextensive statistical mechanics}. This theory was first introduced in 1988 \cite{Tsallis88}, and then refined in 1991 \cite{CuradoTsallis91} and 1998 \cite{TsallisMendesPlastino98}. It is based on the following generalization of $S_{BG}$: 
\begin{equation}
S_{q}\equiv\frac{1-\sum_{i=1}^Wp_i^{\;q}}{q-1}\;\;\;
(q\in{\mathbb R};\;S_1=S_{BG}) \;.
\label{q_entropy}
\end{equation}

Expressions (4), (6) and (7) are respectively generalized into
\begin{equation}
S_q \equiv\frac{1- \int dx \;  [p(x)]^q}{q-1}  \;,
\end{equation}
\begin{equation}
S_q \equiv\frac{1- \int d\Bigl[x/\Bigl(\prod_{r=1}^d \sigma_r \Bigr)\Bigr] \; \Bigl[ \Bigl(\prod_{r=1}^d \sigma_r \Bigr) p(x)\Bigr]^q}{q-1}  \;,
\end{equation}
and
\begin{equation}
S_q \equiv \frac{1-Tr \,\rho^q}{q-1}
\end{equation}

For equiprobability (i.e., $p_i=1/W,\,\forall i$), Eq. (9) yields
\begin{equation}
S_q=\ln_q W \;,
\end{equation}
with the {\it $q$-logarithm} function defined as \cite{Tsallis94}
\begin{equation}
\ln_q z \equiv \frac{z^{1-q}-1}{1-q} \;\;\;(z \in{\mathbb R}; \;z>0; \;\ln_1 z=\ln z) \;.
\end{equation}
Its inverse function, the {\it $q$-exponential}, is given by \cite{Tsallis94}
\begin{equation}
e_q^z \equiv [1+(1-q)z]^{1/(1-q)} \;\;\;(e_1^z=e^z) 
\end{equation}
if the argument $1+(1-q)z$ is positive, and equals zero otherwise.

In Section II we briefly review a few known results concerning the probability distributions which extremize the entropy, and are ultimately associated with macroscopically stationary states. In Section III we briefly review available results related to Fokker-Planck equations. In Section IV we review the standard and the L\'evy-Gnedenko central limit theorems, and argue about the possible formulation of a new theorem that would generalize the standard limit theorem. Such a theorem would provide an important mathematical cornerstone for nonextensive statistical mechanics and its ubiquity in nature. Providing hints that could help formulating and proving the theorem constitutes the main reason of the present paper.

\section{Statistical Mechanics}

{\it Entropy} is necessary to formulate statistical mechanics but it is not sufficient. Indeed, we must also introduce the concept of {\it energy}. The easiest (and more frequently used) way to do so is addressing the so called (by Gibbs) {\it canonical ensemble}. It corresponds to the ubiquitous physical situation in which the system of interest is in contact with a (large by definition) {\it thermostat} which, at equilibrium (or at the physically relevant stationary state, more generally speaking),  imposes to the system its temperature. The system is typically described by a quantum Hamiltonian, and is characterized by the spectrum of energies $\{E_i\} \;(i=1,2,...,W)$ defined as the eigenvalues associated with the Hamiltonian and its boundary conditions. The probability distribution at the relevant stationary state is the one which extremizes the entropy under the norm and the energy constraints. 

\subsection{Boltzmann-Gibbs statistical mechanics}

At thermal equilibrium we must optimize $S_{BG}$ (as given by Eq. (1)) with the norm constraint as given by Eq. (2), and the energy constraint given as follows:
\begin{equation}
\sum_{i=1}^Wp_iE_i=U_{BG} \,,
\end{equation}
where $U_{BG}$ is given and referred to as the {\it internal energy}. By following the Lagrange method, we define the quantity
\begin{equation}
\Phi_{BG} \equiv S_{BG} +   \alpha \Bigl(\sum_{i=1}^W p_i-1\Bigr) - \beta\Bigl( \sum_{i=1}^Wp_iE_i -U_{BG}\Bigr) \,,
\end{equation}
where $\alpha$ and $\beta$ are the Lagrange parameters (their signs have been chosen following tradition). The extremizing condition $\delta \Phi_{BG}/\delta p_j=0$ yields
\begin{equation}
p_j=e^{\alpha -1 - \beta E_j} \;\;\;\;(j=1,2,...,W) \,.
\end{equation}
The use of Eq. (2) allows the elimination of the parameter $\alpha$. We then obtain the celebrated Boltzmann-Gibbs weight
\begin{equation}
p_i=\frac{e^{- \beta E_i}}{Z_{BG}} \;\;\;\;(i=1,2,...,W) \,, 
\end{equation}
where $\beta$ is connected with the thermostat {\it temperature} $T$ through $\beta \equiv 1/T$, and the {\it partition function} is defined as follows:
\begin{equation}
Z_{BG}(\beta) \equiv  \sum_{j=1}^W e^{-\beta E_j}  \,. 
\end{equation}

It is precisely the present Eqs. (19) and (20) that are referred to as {\it dogma} by the mathematician Takens! \cite{Takens91}.

Clearly, if we replace the probability distribution (19) into Eq. (16), we obtain the thermodynamically important relation between the inverse temperature $\beta$ and the internal energy $U_{BG}$. 

As a subsidiary comment, whose relevance will become transparent later on, let us remark that the $BG$ weight (19) can be seen as the solution of the {\it linear} ordinary differential equation
\begin{equation}
\frac{dy}{dx}= ay \;\;\;\;(y(0)=1) \,.
\end{equation}
Indeed, its solution is given by
\begin{equation}
y =e^{ax} \,,
\end{equation}
which reproduces Eq.(19) through the identification $(x,a,y) \equiv(E_i,-\beta,Z_{BG}\,p_i)$. 

\subsection{Nonextensive statistical mechanics}

We want now to optimize $S_q$ (as given by Eq. (9)) with the norm constraint still given by Eq. (2), and the energy constraint generalized as follows (see \cite{TsallisMendesPlastino98}):
\begin{equation}
\frac{\sum_{i=1}^Wp_i^qE_i}{\sum_{i=1}^Wp_i^q}=U_q \,,
\end{equation}
(referred to as {\it $q$-expectation value} or {\it $q$-mean value})
or, equivalently,
\begin{equation}
\sum_{i=1}^Wp_i^q(E_i-U_q)=0 \,.
\end{equation}
The functional (17) is generalized into
\begin{equation}
\Phi_q \equiv S_q +   \alpha \Bigl[\sum_{i=1}^W p_i-1\Bigr] - \beta\Bigl[ \sum_{i=1}^Wp_i^q(E_i -U_q)\Bigr] \,,
\end{equation}
and the extremizing condition $\delta \Phi_q/\delta p_j=0$ yields
\begin{equation}
p_j=\Bigl(\frac{q}{\alpha}\Bigr)^{1/(1-q)}e_q^{- \beta (E_j-U_q)} \;\;\;\;(j=1,2,...,W) \,.
\end{equation}
The use of Eq. (2) allows, as before, the elimination of the parameter $\alpha$, thus obtaining the generalized weight
\begin{equation}
p_i=\frac{e_q^{- \beta (E_i-U_q)}}{Z_q} \;\;\;\;(i=1,2,...,W) \,, 
\end{equation}
with
\begin{equation}
Z_q(\beta) \equiv  \sum_{j=1}^W e_q^{-\beta (E_j-U_q)}  \,. 
\end{equation}
This probability distribution corresponds to a {\it maximum} ({\it minimum}) of $S_q$ for $q>0$ ($q<0$). For $q=0$, the entropy is constant, namely $S_0=W-1$, and the distribution is given by $p_i=[1-\beta(E_i-U_0)] / \sum_{j=1}^W [1-\beta(E_j-U_0)]$ (we recall the {\it cutoff} of the $q$-exponential function for $q<1$, i.e., the states for which $1-\beta(E_i-U_0)<0$ do {\it not} contribute).

As a subsidiary comment, let us remark that the weight (27) can be seen as the solution of the {\it nonlinear} ordinary differential equation
\begin{equation}
\frac{dy}{dx}= ay^q \;\;\;\;(y(0)=1) \,.
\end{equation}
Indeed, the solution is given by
\begin{equation}
y =e_q^{ax} \,,
\end{equation}
which reproduces Eq.(27) through the identification $(x,a,y) \equiv(E_i-U_q \,,-\beta,Z_q\, p_i)$. 

\subsection{The Gaussian case}

Let us now assume a stochastic real variable $x$ such that
\begin{equation}
\langle x \rangle \equiv \int_{- \infty}^{\infty} dx \,x \,p(x)=0 \,,
\end{equation}
and
\begin{equation}
\langle x^2 \rangle \equiv \int_{- \infty}^{\infty} dx \,x^2 p(x)= D \,,
\end{equation}
where $D>0$ is given. The maximization of entropy (4) with constraints (5) and (32) yields the celebrated Gaussian distribution, more precisely
\begin{equation}
p(x) =  \sqrt{\frac{\beta}{\pi}} \, e^{-\beta x^2} \,
\end{equation}
where $\beta$ is a Lagrange parameter. By using Eq. (32), we immediately obtain $\beta=1/2D$, hence
\begin{equation}
p(x) =        \frac{e^{- x^2/(2D)}}{\sqrt{2\pi D}} \,
\end{equation}
We straightforwardly verify that constraint (31) is satisfied as well.

A more general case is to assume the following constraints:
\begin{equation}
\langle x \rangle \equiv \int_{- \infty}^{\infty} dx \,x \, p(x)=C \,,
\end{equation}
and
\begin{equation}
\langle (x-\langle x \rangle)^2 \rangle \equiv \int_{- \infty}^{\infty} dx \,(x- \langle x \rangle)^2 p(x)= D \,.
\end{equation}
The extremizing distribution is then given by
\begin{equation}
p(x) =  \sqrt{\frac{\beta}{\pi}} \, e^{-\beta (x-\langle x \rangle)^2} \,,
\end{equation}
or, equivalently,
\begin{equation}
p(x) =        \frac{e^{- (x- C)^2/(2D)}}{\sqrt{2\pi D}} \,.
\end{equation}

\subsection{The $q$-generalization of the Gaussian case}

Let us assume that the constraints are now even more general, namely
\begin{equation}
\langle x \rangle_q \equiv \frac{\int_{- \infty}^{\infty} dx \,x \,[p(x)]^q}{\int_{- \infty}^{\infty} dx \, [p(x)]^q}=C_q \,,
\end{equation}
and
\begin{equation}
\langle (x-\langle x \rangle_q)^2 \rangle_q \equiv \frac{\int_{- \infty}^{\infty} dx \,(x- \langle x \rangle_q)^2 [p(x)]^q}{\int_{- \infty}^{\infty} dx \, [p(x)]^q}= D_q \,,
\end{equation}
where $P(x) \equiv [p(x)]^q / \int dy\, [p(y)]^q$ is, in the literature, referred to as the {\it escort distribution}.

The associated  distribution extremizing $S_q$ is now given by
\begin{equation}
p(x) =  A_q \sqrt{\beta} \, e_q^{-\beta (x-\langle x \rangle_q)^2}
\,,
\end{equation}
where we have \cite{PratoTsallis99} $A_q =  \sqrt{(q-1)/\pi} \, \Gamma\bigl(1/(q-1)\bigr) / \Gamma\bigl((3-q)/[2/(q-1)]\bigr)        $ for $q>1$, and  $A_q =      \sqrt{(1-q)/\pi} \, \Gamma\bigl((5-3q)/[2(1-q)]\bigr) / \Gamma\bigl((2-q)/(1-q)\bigr)       $ for $q<1$, $\Gamma(z)$ being the Riemann function. 

The use of constraint (40) straightforwardly provides $\beta = 1/[(3-q)D_q]$, which, replaced in Eq. (41), yields
\begin{equation}
p(x) =
\left\{
\begin{array}{ll}   
\frac{A_q}{\sqrt{(3-q) D_q}}  \frac{1}{ [1+   \frac{     q-1}{(3-q)D_q} (x- C_q)^2]^{1/(q-1)}  }   &
(q>1)  \nonumber\\
\frac{A_q}{\sqrt{(3-q) D_q}}  [1-   \frac{     1-q}{(3-q)D_q} (x- C_q)^2]^{1/(1-q)}   &    
(q<1) 
\end{array}
\right.
\end{equation}
In the $q<1$ case, the support is compact, and the distribution vanishes outside the interval $|x-C_q| \le \sqrt{(3-q)D_q/(1-q)}$. 

If $\langle x \rangle_q=0$, distributions (41) and (42) take respectively the simple forms
\begin{equation}
p(x) =  A_q \sqrt{\beta} \, e_q^{-\beta x^2} \,, 
\end{equation}
and
\begin{equation}
p(x) =  \frac{A_q}{\sqrt{(3-q)D_q} \,  e_q^{- x^2/[(3-q)D_q]}} \,,
\end{equation}
which appear frequently in various contexts.

It can be seen \cite{SouzaTsallis97} that these distributions are analytic extensions (to real values of $q$) of the Student's $t$-distribution and the $r$-distribution, for $q>1$ and $q<1$ respectively. Consistently, there is an asymptotic long {\it power-law tail} for $q>1$, and a {\it compact support} for $q<1$. There is an upper bound for $q$, namely $q=3$, imposed by the norm constraint (5). More precisely, the admissible values of $q$ are $q <3$. It deserves to be mentioned that also constraint (40) has an upper bound in order to be finite, which happens to be {\it precisely the same}, i.e. $q=3$.

\section{Diffusion and Fokker-Planck equations}

{\it Normal diffusion} (i.e., the one which satisfies $\langle x^2 \rangle \propto t$, typical of Brownian motion \cite{Einstein05}) is characterized by the heat equation (the simplest form of the Fokker-Planck equation)
\begin{equation}
\frac{\partial p(x,t)}{\partial t}=D\frac{\partial^2 p(x,t)}{\partial x^2} \;\;\;\;(D>0) \,.
\end{equation}
By assuming, at $t=0$, the paradigmatic form
\begin{equation}
p(x,0)=\delta(x) \,,
\end{equation} 
we obtain the following exact solution:
\begin{equation}
p(x,t)= \frac{e^{-x^2/2Dt} }{\sqrt{2 \pi Dt}}\;\;\;\;(t \ge 0)
\end{equation}

There is of course an infinity of possible generalizations of Eq. (45), which is {\it linear} and defined through {\it integer derivatives}. We address here two important such generalizations, both of them associated with {\it anomalous diffusion}, i.e., violating the relation $\langle x^2 \rangle \propto t$ The first one remains linear but replaces the second derivative of the right term by a {\it fractional derivative}; it yields $\langle x^2 \rangle \to\infty \,(\forall t>0)$. The second one is {\it nonlinear} but preserves the integer derivatives; it yields $\langle x^2 \rangle \propto t^\alpha \,(\alpha \ne 1)$.

The first one is as follows:
\begin{equation}
\frac{\partial p(x,t)}{\partial t}=D\frac{\partial^\gamma p(x,t)}{\partial x^\gamma} \;\;\;\;(D>0; 0<\gamma<2) \,.
\end{equation}
The exact solution associated with the $t=0$ condition (46) is given by
\begin{equation}
p(x,t)=   \frac{1}{(Dt)^{1/\gamma}}L_\gamma(x/(Dt)^{1/\gamma})\;\;\;\;(t \ge 0)\,,
\end{equation}
where $L_\gamma(z)$ is the L\'evy distribution of index $\gamma$. The $\gamma \to 2$ limit obviously corresponds to the Gaussian solution (47). The $\gamma=1$ particular case corresponds to the Cauchy-Lorentz distribution
\begin{equation}
p(x,t)= L_1(x/(Dt))=\frac{1}{\pi} \frac{Dt}{(Dt)^2+x^2}
\end{equation}
For all other values of $\gamma$ different from 1 and 2, the L\'evy distribution has no direct analytic expression, and is expressed only through its Fourier transform. 

The second generalization of Eq. (45) is as follows:
 \begin{equation}
\frac{\partial p(x,t)}{\partial t}=D\frac{\partial^2 [p(x,t)]^\nu}{\partial x^2} \;\;\;\;( \nu \in {\mathbb R}) \,.
\end{equation}
The exact solution associated with the $t=0$ condition (46) is given  \cite{PlastinoPlastino95,TsallisBukman96} by
\begin{equation}
p(x,t)= \frac{A_q}{  \sqrt{(3-q)} \,(D t)^{1/(3-q)}  } \,e_q^{-x^2/[(3-q)(Dt)^{2/(3-q)}]    } \;\;\;\;(t \ge 0)
\end{equation}
with
\begin{equation}
q=2-\nu <3 \,,
\end{equation}
with $D>0$ if $\nu>0$, and $D<0$ if $\nu<0$. If we rewrite the diffusion coefficient as  $D \equiv \bar{D}/\nu$, Eq. (51) can be rewritten as follows:
 \begin{equation}
\frac{\partial p(x,t)}{\partial t}=\bar{D}\frac{\partial^2 }{\partial x^2} \frac{[p(x,t)]^\nu-1}{\nu}  \,,
\end{equation}
which, in the limit $\nu \to 0$, becomes
 \begin{equation}
\frac{\partial p(x,t)}{\partial t}=\bar{D}\frac{\partial^2 }{\partial x^2} [\ln p(x,t)]  \,,
\end{equation}
whose solution is known to be the Cauchy-Lorentz distribution.

If we start from any $t=0$ distribution different from (46), the solution $p(x,t)$ does not coincide with Eq. (52), but, nevertheless, it asymptotically approaches \cite{CuradoNobre03} Eq. (52)  for $t\to\infty$. In other words, the solution (52) constitutes an {\it attractor} in the space of the distributions, i.e., it is a solution which is {\it robust}. For all $q<3$, $x$ scales  \cite{TsallisBukman96} with $ t^{1/(3-q)}$. Consequently,  in all cases for which the second moment of $p(x,t)$ is {\it finite} (i.e., for $q<5/3$) we obtain $\langle x^2 \rangle \propto t^{2/(3-q)}$. What happens for $0 \le 5/3<3$ is that the prefactor of $t^{2/(3-q)}$ diverges. In this case, it is convenient to focus on the {\it width} of the distribution, or, equivalently, on $\langle x^2 \rangle_q$. 

To close these remarks about Fokker-Planck equations, let us write down a quite general one, namely
 \begin{equation}
\frac{\partial^\delta p(x,t)}{\partial |t|^\delta}=D\frac{\partial^\gamma [p(x,t)]^{2-q}}{\partial |x|^\gamma} \;\;\;\;( (\delta,\gamma,q) \in {\mathbb R}^3) \,,
\end{equation}
where we have used, for convenience, $q$ rather than $\nu$ (related through Eq. (53)). The regions we address are $0<\delta \le 1$, $0<\gamma \le 2$ and $q<3$. Not surprising, the solutions for arbitrary $(\delta,\gamma,q)$ are not known. But, nevertheless, we depict the $\delta=1$ regions in Fig. 1. This will clarify the particular cases that are related to the central limit theorems we are interested in.
\begin{figure}
\begin{center}
\includegraphics[width=\columnwidth,angle=0]{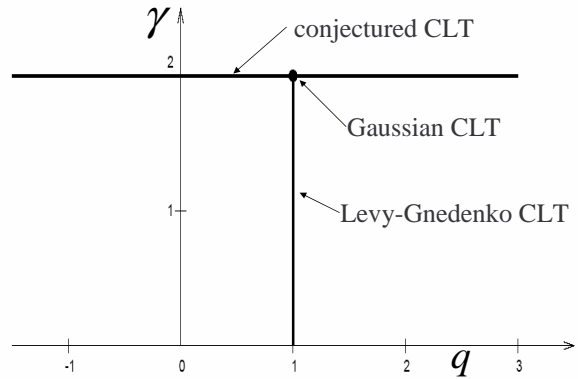}
\vspace{-6.5cm}
\end{center}
\caption{\small $\delta=1$ regions that are relevant for the Central Limit Theorems (CLT). The solution of Eq. (56) for $(\gamma,q)=(2,1)$ is a Gaussian distribution (which optimizes the entropy $S_{BG}$, and whose variance is finite). For $q=1$ and $0<\gamma<2$, the solutions of Eq. (56) are L\'evy distributions (which asymptotically decay as $1/|x|^{1+\gamma}$, and whose variance diverges). For $\gamma =2$ and $q<3$, the solutions are $q$-Gaussians (which optimize the entropy $S_q$, and whose variance is finite for $q<5/3$ and diverges for $5/3 \le q<3$; their $q$-variance is finite for all $q<3$; they have a compact support for $q<1$ and asymptotically decay as $1/|x|^{2/(q-1)}$ for $1<q<3$). Within the stardard CLT, we sum {\it independent} variables whose variance is {\it finite}; within the L\'evy-Gnedenko CLT, we sum {\it independent} variables whose variance {\it diverges}; within the conjectural CLT, we sum {\it specially correlated} variables whose $q$-variance is {\it finite}.
}
\end{figure} 

\section{Central  limit theorems}

In one of its simplest versions, the standard central limit theorem may be formulated as follows. Let $\{x_l\}$ ($l=1,2,...,N$) be a set of {\it independent} random variables, each of them satisfying the {\it same} symmetric (with regard to $x=0$) distribution $p(x)$. Let also $p(x)$ be such that $\langle x^2 \rangle \equiv \int_{-\infty}^{\infty} dx \,x^2 p(x)$ is {\it finite}. Let us define now the sum variable
 \begin{equation}
Z \equiv \sum_{l=1}^N x_l
\end{equation}
The question to which the central limit theorem answers is {\it what is the probability distribution of the random variable $Z$ when $N \to\infty$?} This answer happens to be very simple, namely a {\it Gaussian} (in the properly rescaled variables). If we call $p(Z,N)$ this distribution (with $p(Z,1)=p(x)$), to exhibit the Gaussian we must use as abscissa $Z/\sqrt{N}$, and as ordinate $\sqrt{N}p(Z,N)$. Then, we gradually see emerging the Gaussian as $N\to\infty$. And, {\it what specific Gaussian?} The one whose second moment {\it precisely coincides} with   $\langle x^2 \rangle$.

This theorem clearly is closely related to normal diffusion, hence to Eq. (45) and to its solution (47), where $t$ plays the role of $N$.

Let us now address the other central limit theorem which is known, namely the L\'evy-Gnedenko one. In its simplest version, it can be formulated as follows. As before, the set $\{x_l\}$ ($l=1,2,...,N$) is constituted of {\it independent} random variables, each of them satisfying the {\it same} symmetric (with regard to $x=0$) distribution $p(x)$. But now, we have the case where $\langle x^2 \rangle$ {\it diverges}. The question remains the same, namely {\it what is the probability distribution of the random variable $Z$ when $N \to\infty$?} The answer now is: a {\it L\'evy distribution} $L_\gamma(Z)$. {\it With what value for its index $\gamma$}? All L\'evy distributions decay, for $|x| \to\infty$, as $1/|x|^{1+\gamma}$. This behavior is preserved for the distribution of $Z$ while $N$ increases. Let us be more specific: the typical case is that for which $p(x)$ decays as $1/|x|^\mu$ with $1 < \mu<3$. Then $\gamma=\mu-1$. And {\it what specific L\'evy distribution $L_\gamma$?} The one which has {\it precisely the same coefficient} of the asymptotically dominant term. In other words, let us consider the {\it finite} value $\lim_{|x|\to\infty}[p(x) |x|^\mu]$. The particular L\'evy distribution which is asymptotically approached when $N \to\infty$ is the one which has the {\it same} value for $\lim_{|x|\to\infty}[L_\gamma(x) |x|^{1+\gamma}]$. To see, on a graphic representation, the gradual emergence of the L\'evy distribution while $N$ increases, the abscissa must be $Z/N^{1/\gamma}$, and the ordinate must be $N^{1/\gamma}p(Z,N)$.

This theorem is closely related to the anomalous diffusion characterized by Eq. (48) and by its solution (49), where, as before, $t$ plays the role of $N$.

Let us finally address the {\it conjectural} theorem that we are focusing in this paper. It is of course the one to be associated with the nonlinear Eq. (51) and its solution (52). {\it What hypothesis is to be violated in the two preceding and well studied theorems?} {\it The hypothesis of independency!} Indeed, we believe that the variables $\{x_l\}$ are to be assumed somehow {\it collectively correlated} in such a persistent manner that the correlation does {\it not} disappear even in the $N \to\infty$ limit. For the standard central limit theorem, the quantity which is preserved is the second moment $\langle x^2\rangle \equiv \int_{-\infty}^{\infty} dx\,x^2 p(x)$. For the L\'evy-Gnedenko theorem, the quantity which is preserved is the coefficient $\lim_{|x|\to\infty}[p(x) |x|^\mu]$. For this conjectural theorem, some quantity is expected to be preserved. Could it be $\langle x^2 \rangle_q \equiv \int_{-\infty}^{\infty}dx \,x^2 [p(x)]^q /  \int_{-\infty}^{\infty}dx \, [p(x)]^q$ ?

It should be transparently clear at this point that we have {\it no definitive arguments} for proving this conjectural theorem. Various converging paths are nevertheless available that might inspire a (professional or {\it amateur}) mathematician the way to prove it. Galileo used to say that knowing a result is not neglectable in order to prove it! It is our best hope that his saying does apply in the present case! {\it So, what are these converging paths?} Although naturally intertwingled, let us expose them along six different lines.

\subsubsection{The $q$-product hint}

The following generalization of the product operation has been recently introduced \cite{Borges04}. It is called {\it $q$-product} and is defined through
\begin{equation}
X \otimes_q Y \equiv [X^{1-q}+Y^{1-q}-1]^{1/(1-q)} \;\;\;\;(q \in {\mathbb R})
\end{equation}
We shall address the case where $X \ge 1$ and $Y \ge 1$ \cite{remark}. This product has the following properties:

(i) $X \otimes_1 Y=XY $;

(ii) $\ln_q (X \otimes_q Y)=\ln_q X+ \ln_q Y$ (whereas $\ln_q (X Y)=\ln_q X+ \ln_q Y+(1-q)(\ln_q X)( \ln_q Y)$; 

(iii) $1/(X \otimes_q Y)=(1/X) \otimes_{2-q}(1/Y)$;

 (iv) $X \otimes_q(Y \otimes_q Z)=(X \otimes_q Y) \otimes_q Z=X \otimes_q Y \otimes_q Z= (X^{1-q}+Y^{1-q}+Z^{1-q}-2)^{1/(1-q)}$; 

(v) $X \otimes_q 1=X$;

(vi) $X \otimes_q Y =Y \otimes_q X$;

(vii) For fixed $(X,Y)$, $X \otimes_q Y$ monotonically increases with $q$ .

Property (ii) is particularly important since it is directly relevant to the {\it extensivity} of $S_q$ that will be addressed soon.

Notice that property (iii) involves, like the solution of Eq. (51) with Eq. (53), a $q \leftrightarrow (2-q)$ transformation. 
 
We may apply this product to the number $W_{A_1+A_2+...+A_N}^{\mbox{\it eff}}$ of {\it allowed} states in a composed system whose subsystems $A_1,A_2,...,A_N$ have respectively $W_{A_1}, W_{A_2},...,W_{A_N}$ possible states (by {\it allowed} we mean that their probability is essentially {\it nonzero}). If $q=1$ we have the total number $W_{A_1+A_2+...+A_N}=\prod_{l=1}^N W_{A_l}$ of states that are not only possible a priori, but even generically allowed. But, if $q \ne 1$, say $q<1$, we expect correlations to inhibit (even forbidden occasionaly) {\it some} of the states, i.e., to be associated with a probability close (in some sense) to zero. In this case, the {\it effective} total number $W_{A_1+A_2+...+A_N}^{\mbox{\it eff}}$ of allowed states is expected to be {\it smaller} that  $W_{A_1+A_2+...+A_N}$. We expect to have basically 
\begin{eqnarray}
W_{A_1+A_2+...+A_N}^{\mbox{\it eff}} &\sim& W_{A_1} \otimes_q W_{A_2} \otimes_q ...\otimes_qW_{A_N}  \nonumber \\
&=&\Bigl[ \Bigl(\sum_{l=1}^N W_{A_l}^{(1-q)}\Bigr) 
-(N-1)\Bigr]^{1/(1-q)}  \nonumber \\
&<& W_{A_1+A_2+...+A_N}=\prod_{l=1}^N W_{A_l} \, . 
\end{eqnarray}
In particular, for $q=0$, we have 
\begin{equation}
W_{A_1+A_2+...+A_N}^{\mbox{\it eff}}= \Bigl(\sum_{l=1}^N W_{A_l}\Bigr) -N+1 \,.
\end{equation}
If the $N$ subsystems are all equal, we have that
\begin{equation}
\ln_q W^{\mbox{\it eff}}(N) \sim N \ln_q W(1),
\end{equation} 
the changement of notation clearly being $W^{\mbox{\it eff}}(N) \equiv W_{A_1+A_2+...+A_N}^{\mbox{\it eff}}$ and $W(1) \equiv W_{A_1}$. 
Consequently, for $q=1$, we have 
\begin{equation}
W^{\mbox{\it eff}}(N)=W(N)=[W(1)]^N \,, 
\end{equation}
whereas, for $ q < 1$, we have 
\begin{eqnarray}
W^{\mbox{\it eff}}(N) &\sim& \Bigl\{N\Bigl[[W(1)]^{1-q}-1\Bigr]+1\Bigr \}^{1/(1-q)} \nonumber \\
&<& W(N)=[W(1)]^N \,. 
\end{eqnarray}
We therefore see that, for $q=1$, the number $W^{\mbox{\it eff}}(N)$ of nonzero-probability states grows {\it exponentially} with $N$, whereas, for $q<1$, $W^{\mbox{\it eff}}(N)$ grows like a {\it power-law} with $N$, more precisely like $N^{1/(1-q)}$. For $q=0$, it grows {\it linearly} with $N$, more precisely $W^{\mbox{\it eff}}(N)=N[W(1)-1]+1$. 

\subsubsection{The hint of the extensivity of $S_q$}

We have recently argued (see \cite{SatoTsallis04} and references therein) that special correlations may exist between the $N$ subsystems $A_1, A_2,...,A_N$ of a composed system such that one (and only one) value of the index $q$ exists which ensures the {\it additivity} of $S_q$. The trivial case of course is that of $S_{BG}$. We consider the case of {\it independency}, i.e., the joint probabilities given by
\begin{equation}
p_{i_1i_2...i_N}^{A_1+A_2+...+A_N}=\prod_{l=1}^N p_{i_l}^{A_l} \;\;,\forall (i_1,i_2,...,i_N).
\end{equation}
We straightforwardly verify that
\begin{equation}
S_{BG}(A_1+A_2+...+A_N)=\sum_{l=1}^N S_{BG}(A_l)  
\end{equation}
where
\begin{eqnarray}
&&S_{BG}(A_1+A_2+...+A_N)   \equiv  \\ 
&&-\sum_{i_1=1}^{W_{A_1}}\sum_{i_2=1}^{W_{A_2}}  ... \sum_{i_N=1}^{W_{A_N}}  p_{i_1i_2...i_N}^{A_1+A_2+...+A_N} \ln p_{i_1i_2...i_N}^{A_1+A_2+...+A_N}  \nonumber
\end{eqnarray}
and
\begin{equation}
S_{BG}(A_l) \equiv -  \sum_{i_l=1}^{W_{A_l}} p_{i_l}^{A_l} \ln p_{i_l}^{A_l} \,,\forall l.
\end{equation}

We can also verify
\begin{equation}
\ln[1+(1-q)S_q(A_1+A_2+...+A_N)]=\sum_{l=1}^N \ln[1+(1-q)S_q(A_l)] \;. 
\end{equation}

If the subsystems are all equal, we have
\begin{equation}
S_{BG}(N)=N S_{BG}(1) \,,  
\end{equation}
the notation being self-explanatory. 

If we consider $N=2$ in Eq. (68) we obtain
\begin{equation}
S_{q}(A_1+A_2)=S_{q}(A_1) +S_{q}(A_2)+(1-q)S_q(A_1)S_q(A_2) \;.
\end{equation}
Therefore, $S_{BG}$ is said to be {\it extensive} (or {\it additive}). Consistently, $S_q$ is, unless $q=1$, {\it nonextensive} (or {\it nonadditive}). It is in fact from this property that the statistical mechanics we are talking about has been named {\it nonextensive}. As we shall exhibit in what follows, this early denomination might (unfortunately) be somewhat misleading. Indeed, $S_q$ is, for $q \ne 1$, nonextensive {\it if} the subsystems are (explicitly or tacitly) assumed {\it independent}. But, if they are specially correlated, $S_q$ can in fact be extensive for some special value of $q \ne 1$.

Consider now the following set of joint probabilities (clearly corresponding to {\it nonindependent} subsystems):
\begin{equation}
p_{11...1}^{A_1+A_2+...+A_N}=\Bigl(\sum_{l=1}^N p_1^{A_l}\Bigr) - N+1 \,,
\end{equation}
\begin{equation}
p_{1...l_i...1}^{A_1+A_2+...+A_N}=p_{i_l}^{A_l} \;\;\;(i_l \ne 1; \forall l) \,,
\end{equation}
and zero otherwise. Consequently, we have only $ W_{A_1+A_2+...+A_N}^{\mbox{\it eff}}= \Bigl(\sum_{l=1}^N W_{A_l}\Bigr) -N+1$  joint probabilities which are generically nonzero, consistently with Eq. (60). We straightforwardly verify 
\begin{equation}
S_0(A_1+A_2+...+A_N)=\sum_{l=1}^N S_0(A_l)  \,,
\end{equation}
where
\begin{equation}
S_0(A_1+A_2+...+A_N)=W_{A_1+A_2+...+A_N}^{\mbox{\it eff}}-1 \,,
\end{equation}
and
\begin{equation}
S_0(A_l) = W_{A_l}-1 \;\;\;(\forall l) \,.
\end{equation}
Therefore, for the particular correlations involved in Eqs. (71) and (72), it is for $q=0$ (and {\it not} for $q=1$) that $S_q$ is additive. 

Unfortunately, we do not  know the general explicit form of the set of joint probabilities for an arbitrary number $N$ of (not necessarily equal) subsystems, corresponding to $q \ne 0,1$. We do know however, for $0 \le q \le 1$, some special cases, such as the $N=2, 3$ generic {\it binary} subsystems. They are presented and analyzed in \cite{SatoTsallis04}. 
We also know, for binary equal subsystems, exact recurrence relations that enable the calculation of the entire set of $2^N$ joint probabilities associated with $N$ subsystems given the entire set of $2^{N-1}$ joint probabilities associated with $N-1$ subsystems. This calculation is rather lengthy and will be the object of a separate paper. 

Now we present here (in a form which is slightly different, and possibly more transparent, than the one used in \cite{SatoTsallis04}) the case of $N=2$ equal binary subsystems. Consider the set of joint probabilities indicated in Table I. Let us impose the additivity of the entropy, i.e., $S_q(2)=2S_q(1)$, where
\begin{equation}
S_q(2) \equiv \frac{1- [p^2 +\kappa]^q-2[p\,(1-p)-\kappa]^q-[(1-p)^2+\kappa]^q}{q-1}
\end{equation} 
and
\begin{equation}
S_q(1) \equiv \frac{1- p^q-(1-p)^q}{q-1} \,.
\end{equation} 
The relation between the {\it correlation} $\kappa$, $p$ and $q$ is then given by
\begin{equation}
2   [p^q+(1-p)^q] - [p^2 +\kappa]^q-2[p\,(1-p)-\kappa]^q-[(1-p)^2+\kappa]^q = 1        \,. 
\end{equation} 
(With the notation $f_q(p) \equiv p^2 + \kappa$, this relation coincides with the one indicated in \cite{SatoTsallis04}). In Fig. 2, typical $\kappa$ versus $p$ curves are presented. The {\it lower} curve corresponds to $p^2+\kappa=0$ and to $(1-p)^2+\kappa=0$. The {\it upper} curve corresponds to $p(1-p)-\kappa=0$. The lower curve undoubtedly corresponds to $q=0$. The upper curve  could in principle correspond to {\it both} cases $q \to -\infty$ and $q \to \infty$. Indeed, for $\kappa=p(1-p)$, we have that $p^2+\kappa=p$ and $(1-p)^2+\kappa=1-p$, hence $S_q(2)=S_q(1)$. Since we must also satisfy  $S_q(2)=2S_q(1)$, only two possibilities emerge a priori, and these are $S_q(1)=0$ (which corresponds to $q \to\infty$), and $S_q(1)\to\infty$ (which corresponds to $q \to -\infty$). Monotonicity with regard to $q$ suggests that the upper curve should correspond to $q\to\infty$. But on the other hand, having, among the four joint probabilities that are a priori possible in the present $N=2$ system, {\it two} zeros  (which definitively is what corresponds to the upper curve) appears as {\it more} restrictive than having {\it one} zero (which definitively is what corresponds to the lower curve). The generic {\it four}  nonzero-probability case corresponds to $q=1$, and the generic {\it three} nonzero-probability case corresponds to $q=0$. Consequently, it appears as reasonable that the generic {\it two} nonzero-probability case would correspond to $q \to -\infty$. This kind of paradoxal situation appears to need further discussion to be clarified. The situation would probably be clear-cut if we had --- but we do not! ---, for arbitrary values of $N$ arbitrary subsystems, the general answer to be associated with arbitrary $q$. This point thus remains open.
\begin{table}[htbp]
\begin{center}
\begin{tabular}{c||c|c||c}
$_A\setminus^B$    &  1                                    & 2                                                                   \\[1mm] \hline\hline
1  &  $\;\;p^2 +\kappa\;\;$                                  & $\;\;p\,(1-p)-\kappa\;\;$              & $\;\;p\;\;$   \\[3mm] \hline
2  &  $p\,(1-p)-\kappa    $                                  & $(1-p)^2+\kappa$                      & $1-p$       \\[3mm] \hline \hline
    &  $p$                                                            & $1-p$                                        & 1

\end{tabular}
\end{center}
\caption{Joint probabilities for two binary subsystems $A$ and $B$. The marginal probabilities are indicated as well. The correlation $\kappa$ and the probability $p$ are such that all terms $p^2 +\kappa$, $p\,(1-p)-\kappa$, and $(1-p)^2+\kappa$ remain within the interval $[0,1]$. The case of independency corresponds to $\kappa=0$. 
}
\end{table}
\begin{figure}
\begin{center}
\vspace{-1.5cm}
\includegraphics[width=\columnwidth,angle=0]{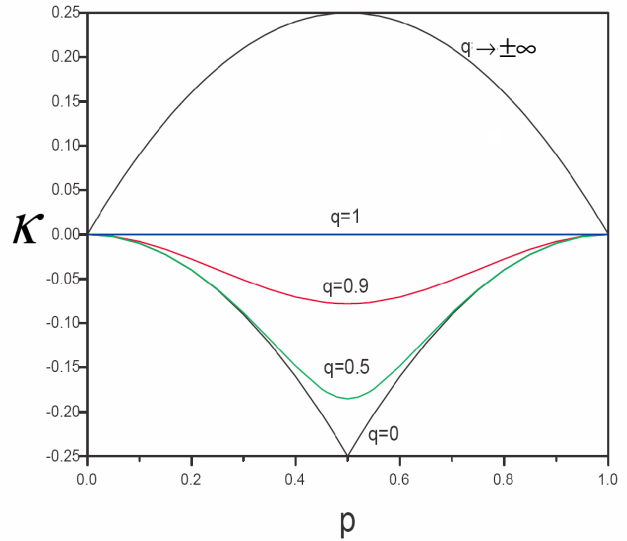}
\vspace{-3.2cm}
\end{center}
\caption{\small Curves $\kappa(p)$ which, for typical values of $q$, imply additivity of $S_q$. For $ -1/4 \le \kappa \le 0$ we have $\sqrt{-\kappa} \le p \le 1- \sqrt{-\kappa}$. For $0 \le \kappa \le 1/4$ we have $(1-\sqrt{1-4\kappa})/2 \le p \le (1+\sqrt{1-4\kappa})/2$ .
}
\end{figure} 

It is worthy stressing a logical consequence of what has been said up to now. Unless $q=1$, we {\it cannot} simultaneously have equal probabilities for the set of joint probabilities of the {\it allowed} states {\it and} for the associated marginal probabilities. For example, for the $q=0$ case of two equal binary subsystems, equal probabilities for the joint set means $p_{11}^{A+B}=p_{12}^{A+B}=p_{21}^{A+B}=1/3$ and $p_{22}^{A+B}=0$, hence $p=2/3$ (which differs from $ 1-p=1/3$), whereas equal probabilities for the marginal sets means $p=1/2$, hence $p_{11}^{A+B}=0 \ne p_{12}^{A+B}=p_{21}^{A+B}=1/2$. The $q=1$ case {\it simultaneously} allows for $p_{11}^{A+B}=p_{12}^{A+B}=p_{21}^{A+B}=p_{22}^{A+B}=1/4$ and $p=1-p=1/2$.  It is clear that, if such behavior (i.e., generic impossibility of equal probabilities for {\it both} joint and marginal ones simultaneously) persists up to the thermodynamic limit $N \to\infty$, it will have heavy consequences for the macroscopic statistics of the system.

Let us recall at this stage that, depending on whether the subsystems that we are composing are or are not independent, it is $S_{BG}$ or a different entropy which is {\it additive}. If they are correlated in the special form that has been illustrated above, and that persists up to the thermodynamic limit $N \to\infty$, it is $S_q$ with a specific value of $q \ne 1$ which becomes {\it extensive} \cite{remark2}. The corresponding statistical mechanics should consistently be based on $S_q$ (or on a directly related one, such as say $S_{2-q}$) and {\it not} anymore on $S_{BG}$. 
We may summarize this scenario through the following statement: {\it Unless the composition law of the subsystems is specified, the question whether an entropy (or some similar quantity) is or is not extensive has no sense}. Allow us a quick digression. The situation is totally analogous to the quick or slow motion of a body. Ancient greeks considered motion to be an absolute property. It was not until Galileo that it was clearly perceived that motion has no sense {\it unless the referential is specified}.   

\subsubsection{The hint of the $q$-generalization of the Pascal triangle}

Very recently, Suyari and Tsukuda \cite{SuyariTsukuda04} used the concept of $q$-product to consistently generalize various results that are widely known in the context of $BG$ statistical mechanics and its mathematical structure. The generalization of the $n!$ ($n$-factorial) operation, as well as of the binomial (and even multinomial) coefficients was performed. As a corollary, the Pascal triangle itself was $q$-generalized as well. It is known that the coefficients of the $n$-th line of the Pascal triangle yield (after appropriate centralization and rescaling) the Gaussian distribution in the $n \to \infty$ limit. This is well known to be a simple consequence of the standard central limit theorem as applied to independent binary variables. Suyari claims that, in the $n \to \infty$ limit of the $q$-generalized Pascal triangle, what is obtained is precisely the $q$-Gaussian distributions!

\subsubsection{The hint of the scale-free networks}

The mathematical study of {\it random networks} (or {\it random graphs}) started many decades ago. But quite recently, it has acquired great interest due to the fact that such structures appear to be ubiquitous in physical, social, internet and other complex phenomena (see \cite{WattsStrogatz98,AlbertBarabasi02} and references therein). A central quantity of such structures is the so-called {\it connectivity} or {\it degree distribution}, defined as the probability distribution of the number of links that are connected to the same site (or node); being more explicit, what one counts is the percentage of nodes that have a given number of links. For the important class of networks that are referred to as {\it scale-free} ones, this distribution is systematically found to be precisely a $q$-exponential. Many examples do exist in the literature. Three recent illustrations can be found in \cite{AlbertBarabasi00,Doye02,SoaresTsallisMarizRodrigues04}. 

These scale-invariant structures exhibit {\it hubs} and {\it sub-hubs}, and so on. If embedded into a $d$-dimensional space, they tend to have {\it zero} Lebesgue measure. 
They appear to be like fractals, which implies that most of the locations of the $d$-dimensional space are forbidden (like the flights of any air company are expected to start and end {\it only} at the airports where that company operates, and not in any place of the territory!).   

\subsubsection{The nonlinear dynamical hint}

We review at this point a few nonlinear dynamical results which, although not directly related to the possible {\it third} central limit theorem we are seeking, provide what we consider important connections within the mathematical structure which sustains nonextensive statistical mechanics. 

We focus on classical nonlinear dynamical systems, either conservative (e.g., Hamiltonian systems) or dissipative. The basic ideas are quite general, but their presentation becomes easier if we illustrate them on a simple, one-dimensional, system. Let us address unimodal one-dimensional maps, such as for example the $z$-logistic map defined through 
\begin{eqnarray}
x_{t+1}&=&1-a |x_t|^z     \nonumber \\
(t=0,1,2,...&;& z \ge 1; \,0 \le a \le 2;\, -1 \le x_t \le 1) 
\end{eqnarray}
The {\it sensitivity $\xi$ to the initial conditions} is defined as follows:
\begin{equation}
\xi(t) \equiv lim_{\Delta x(0) \to 0} \frac{\Delta x(t)}{\Delta x(0)} \,,
\end{equation}
where $\Delta x(t)$ is the discrepancy at time $t$ of two values of $x_0$ initially separated by $\Delta x(0)$. The typical behavior of $\xi$ is given by
\begin{equation}
\xi(t)= e^{\lambda_1 t} \,,
\end{equation} 
where $\lambda_1$ is referred to as the {\it Lyapunov exponent} (the subindex 1 will become clear in a few lines). The following ordinary differential equation is satisfied: $d\xi/dt=\lambda_1\xi$. 

For most values of the {\it control parameter} $a$ (see Eq. (79)), $\lambda_1$ is nonzero. When it is positive, we shall say that there is {\it strong chaos}. When it is negative, we have {\it regular orbits} as attractors (e.g., fixed points, cycles-2, cycles-3, cycles-4, etc). But there is an infinity of values of $a$ for which $\lambda_1$ vanishes. Examples are (i) all the values of $a$ for which there is a {\it doubling-period bifurcation} from one cycle to its double; (ii) all the values of $a$ for which there is a {\it tangent bifurcation}; (iii) all the values of $a$ for which there is an {\it edge to chaos}. All these cases are interesting, but by far the richest one is case (iii), referred to as {\it weak chaos}. Indeed, it is there that we expect the appearence of complexity (in physics, biology, economics, linguistics, and elsewhere), since it is the frontier between considerable order (regular orbits) and considerable disorder (strong chaos). So, the following question arises: {\it what is the behavior of $\xi(t)$ when $\lambda_1$ vanishes?}  The typical behavior is as follows:
\begin{equation}
\xi(t)= e_q^{\lambda_q t} \,,
\end{equation} 
where we have $q>1$ (with $\lambda_q>0$) for the above cases (i) and (ii), and $q<1$ (with $\lambda_q>0$) for the above case (iii). The following ordinary differential equation is satisfied: $d\xi/dt=\lambda_q\, \xi^q$. The behavior (82) was conjectured in 1997 \cite{TsallisPlastinoZheng97} and was proved in 2003 \cite{BaldovinRobledo03}. For example, for $z=2$, the first edge of chaos appears at $a=a_c \equiv 1.4011...$. Its associated values are $q_{sen}=0.2445...$ and $\lambda_{q_{sen}}=1/(1-q_{sen})=1.3236...$, where the subscript {\it sen} stands for {\it sensitivity}.

Let us now address another very important quantity, namely the {\it entropy production $K$ per unit time}. It is basically defined as follows. Take the admissible phase space (the interval $[-1,1]$ for the variable $x$ in the map (79), for instance), and make a partition of it in $W$ little parts, identified through $i=1,2,...,W$. Then consider $N_0$ initial values of $x_0$ within one of the $W$ intervals. As a function of time, the points within the initial window will possibly spread around, in such a way that we have $N_i(t)$ points in the $i$-th interval ($\sum_{i=1}^W N_i(t)=N_0$). We define next the set of probabilities $p_i \equiv N_i(t)/N_0\,(\forall i)$, and finally we calculate the entropy $S_q(t)$ through Eq. (9) for any chosen value of $q$. By construction we have $S_q(0)=0$. We then define the entropy production per unit time as follows 
\begin{equation}
K_q \equiv \lim_{t \to\infty} \lim_{W \to \infty} \lim_{N_0 \to \infty} \frac{S_q(t)}{t} \,.
\end{equation}
The rate $K_1$ coincides, in most cases, with the so called Kolmogorov-Sinai entropy rate. The rate $K_q$ represents its $q$-generalization. Consistently with the Pesin theorem, we expect, whenever $\lambda_1 \ge 0$,
 \begin{equation}
K_1=\lambda_1 \,.
\end{equation}
This is indeed verified (see, for instance, \cite{LatoraBaranger99}). For example, for $(z,a)=(2,2)$ in (79), we obtain $K_1=\lambda_1= \ln 2$. 
What happens however at the edge of chaos (say at $a_c(z)$)? Can we state something more informative than just $K_1=\lambda_1=0$? {\it Yes, we can.} It can be numerically shown \cite{LatoraBarangerRapisardaTsallis00} and analytically proved \cite{BaldovinRobledo04} that
 \begin{equation}
K_{q_{sen}}=\lambda_{q_{sen}} \,.
\end{equation}
This interesting relation (which refers in fact to {\it upper bounds}: see details in \cite{BaldovinRobledo04}) basically generalizes the celebrated Pesin theorem, and was also  conjectured in 1997 \cite{TsallisPlastinoZheng97}. To be more specific, what happens is that there is a value of $q$ (and only one), noted $q_{sen}$, such that $K_q =0$ for $q>q_{sen}$, and $K_q \to\infty$ for $q<q_{sen}$. And for $q=q_{sen}$ we obtain a finite value $K_{q_{sen}}$ which precisely coincides with $\lambda_{q_{sen}}$. When $\lambda_1>0$, $q_{sen}=1$ (strong chaos); when  $\lambda_1=0$ and the system is at the edge of chaos, $q_{sen}<1$ and generically $\lambda_{q_{sen}}>0$ (weak chaos). For example, for the edge of chaos of the universality class to which the map (79) belongs, $q_{sen}$ increases from $-\infty$ to slightly below unity when $z$ increases from 1 to $\infty$.

What happens when our nonlinear dynamical system has more than one, say $\mu$ nonnegative Lyapunov exponents $\{\lambda_1^{(m)}\}$ ($m=1,2,...,\mu$)? From the Pesin theorem, we certainly expect
 \begin{equation}
K_1=\sum_{m=1}^\mu \lambda_1^{(m)} \,.
\end{equation} 
What happens then if the system is at an edge of chaos, where $\lambda_1^{(m)}=0 \,,\forall m$? We expect \cite{AnanosBaldovinTsallis04} the following (conjectural) behavior for the Lebesgue measure $\xi$ associated with the dynamically expanding directions:
\begin{equation}
\xi \simeq \prod_{m=1}^\mu \xi^{(m)} =  \prod_{m=1}^\mu exp_{q_{sen}^{(m)}} \bigl( \lambda_{q_{sen}^{(m)}}^{(m)} \, t \bigr) \,,
\end{equation}
where the $(\simeq)$ sign might become just $(=)$ under some simplifying hypothesis (like orthogonality of the directions along which the expansions associated with the $\lambda_1^{(m)}$'s occur). 

Since we essentially expect, for $t \to\infty$,  
\begin{equation}
S_{q_e}\sim \ln_{q_e} \xi \,,
\end{equation}
(the subscript $e$ stands for {\it entropy}), we possibly have the following relation:
\begin{equation}
S_{q_e}\sim \ln_{q_e}        \Bigl[   \;       \prod_{m=1}^\mu exp_{q_{sen}^{(m)}} \bigl( \lambda_{q_{sen}^{(m)}}^{(m)} \, t \bigr)   \Bigr]             \,.
\end{equation}
Using the definition (83) we obtain the following interesting relation:
\begin{equation}
K_{q_e} = \lim_{t \to\infty} \frac{ \ln_{q_e}       \Bigl[   \;       \prod_{m=1}^\mu exp_{q_{sen}^{(m)}} \bigl( \lambda_{q_{sen}^{(m)}}^{(m)} \, t \bigr)   \Bigr] }{t}            \,.
\end{equation}
Two important cases must be distinguished, namely strong and weak chaos. Strong chaos corresponds to $\lambda_1^{(m)}>0 \,,\forall m$. In this case, we have $q_e=q_{sen}^{(m)}=1\,,\forall m$, hence Eq. (90) straightforwardly recovers relation (86), consistently with the Pesin theorem. Weak chaos corresponds to $\lambda_1^{(m)}=0 \,,\forall m$. In this case, by focusing on the $t \to\infty$ asymptotic  region, we have 
 \begin{eqnarray}
K_{q_e} = \frac{       \prod_{m=1}^\mu  \Bigl[   \bigl(1-q^{(m)}_{sen} \bigr) \lambda^{(m)}_{q^{(m)}_{sen}}  \Bigr]^{\frac{1-q_e}{1-q^{(m)}_{sen}}}      }{1-q_e}     \nonumber \\ 
\times      \lim_{t \to\infty} \frac{      t^{(1-q_e)\sum_{m=1}^\mu                    \frac{1}{1-q^{(m)}_{sen}}}                 }{t} \,,
\end{eqnarray}
consequently (since $q_e$ must be chosen so that $K_{q_e}$ is {\it finite})
\begin{equation}
\frac{1}{1-q_e}  = \sum_{m=1}^\mu \frac{1}{1-q_{sen}^{(m)}} \,,
\end{equation}
and
\begin{equation}
K_{q_e} = \frac{       \prod_{m=1}^\mu  \Bigl[   \bigl(1-q^{(m)}_{sen} \bigr) \lambda^{(m)}_{q^{(m)}_{sen}}  \Bigr]^{\frac{1-q_e}{1-q^{(m)}_{sen}}}      }{1-q_e} \,.
\end{equation}
Eq. (93) can be rewriten in a more symmetric form, namely
\begin{equation}
[(1-q_e)K_{q_e}]^{\frac{1}{1-q_e}} =     \prod_{m=1}^\mu  \Bigl[   \bigl(1-q^{(m)}_{sen} \bigr) \lambda^{(m)}_{q^{(m)}_{sen}}  \Bigr]^{\frac{1}{1-q^{(m)}_{sen}}}       \,.
\end{equation}

If $\mu=1$, Eq. (92) recovers $q_e=q_{sen}$, and Eq. (94) recovers Eq. (85). 

Another interesting particular case is when $q_{sen}^{(m)}=q_{sen} \,(\forall m)$,  and $\lambda^{(m)}_{q^{(m)}_{sen}} =  \lambda_{q_{sen}} \,(\forall m)$. From Eqs. (92) and (94) it follows then
\begin{equation}
1-q_e=\frac{1-q_{sen}}{\mu}
\end{equation}
and
\begin{equation}
K_{q_e}=\mu  \, \lambda_{q_{sen}} \,.
\end{equation}
An example which verifies Eq. (95) can be found in \cite{AnanosBaldovinTsallis04}.

One more particular case which is interesting is when the expansion is {\it linear} in time for {\it all} directions, i.e., $q_{sen}^{(m)} = 0 \, (\forall m)$. We then have, from Eqs. (92) and (94), 
\begin{equation}
q_e=1-\frac{1}{\mu} 
\end{equation}
and
\begin{equation}
K_{q_e}=\mu  \, \Bigl[\prod_{m=1}^\mu        \lambda_0^{(m)} \Bigr]^{1/\mu} \,.
\end{equation}
Relation (97) has already emerged in the literature through various forms. A first example concerns $\mu=1$, hence we expect $q_e=0$. This is precisely what can be verified \cite{CasatiTsallisBaldovin04} for the Casati-Prosen triangle map \cite{CasatiProsen}, which is a two-dimensional, conservative (hence $\mu=1$), mixing, ergodic one, with {\it linear} instability. In addition to $q_e=0$, it has been verified that $K_e=\lambda_0$, in accordance with Eq. (98). A second example is the $\mu$-dimensional lattice Lotka-Volterra, for which it has precisely been verified \cite{TsekourasProvataTsallis04,Anteneodo04} the result (97).  A third example is a specific  $d$-dimensional Boltzmann latice model \cite{BoghosianLoveCoveneyKarlinSucciYepez03}. Its Hamiltonian-like behavior leads to $\mu=d/2$, which, from Eq. (97), implies $q_e=1-2/d$. It is precisely this result that the authors \cite{BoghosianLoveCoveneyKarlinSucciYepez03} have obtained by imposing Galilean invariance to the dynamical equations of their model.

It is clear that the interesting (and possibly exact) relations (92) and (94) remain to be proved. At the present stage they constitute but conjectures.  

\subsubsection{The empirical hint}

Last but not least, let us present a very pragmatic, epistemological-like,  reason. There are, in the literature, already so many  natural and artificial systems (and their number constantly increases) whose central quantities are well fitted  by $q$-exponentials and $q$-Gaussians, that one is compelled to believe that only a limit theorem could explain such an ubiquity.

\section{Conclusions}

In this final section, let us summarize the scenario within which we are working. It appears that the entropic form $S(N; \{p_i\})$ to be used for constructing a statistical mechanics that is naturally compatible with usual macroscopic thermodynamics should be {\it generically extensive}, i.e., such that 
\begin{equation}
0 \le \lim_{N \to \infty}\frac{S(N;\{p_i\})}{N}< \infty \,,
\end{equation}
the equality occuring only for the case of certainty or (thermodynamically) close to it (the typical example is a system in themal equilibrium at {\it zero temperature}).  Within this (axiomatic) viewpoint, we can distinguish the following cases:

(i) The $N$ subsystems (typically physical elements such as particles free to move translationally and/or rotationally, localized spins, and similar entities) are (probabilistically) {\it independent} (the paradigmatic case is that of ideal gases; their total energy is strictly proportional to $N$, i.e., additive, hence {\it extensive}). Then we must use $S_{BG}$ since 
$S_{BG}(N; \{p_i\}) \propto N \;(\forall \{p_i\})$.

(ii) The $N$ subsystems are {\it locally correlated} (the paradigmatic case is that of {\it short}-range-interacting many-body Hamiltonian systems; their total energy is asymptotically proportional to $N$, i.e., it is {\it extensive} once again). Then, once again, we must use $S_{BG}$ since it satifies Eq. (99), $\forall \{p_i\}$.

(iii) The $N$ subsystems are {\it globally correlated} in the special manner addressed in this paper (the paradigmatic case appears to be that of {\it long}-range-interacting many-body Hamiltonian systems; their total energy asymptotically increases with $N$ faster than linearly, i.e., it is {\it nonextensive}). Then, we must use $S_q$ for that unique value of $q$ which guarantees Eq. (99), $\forall \{p_i\}$.

(iv) The $N$ subsystems are {\it globally correlated} in a manner which is more complex (or just as complex but in a different manner) than the one addressed in this paper. Then we must use an entropy which is {\it not} included in the family $S_q$ for any value of $q$. Such an entropy would have to be either more general than $S_q$ (see for instance \cite{TsallisSouza2003} for the so called {\it superstatistics} \cite{BeckCohen2003}), or just of a completely different type (see for instance \cite{Curado99,AnteneodoPlastino99,KaniadakisLissiaScarfone04}).

The cases (i) and (ii), which we may call {\it simple} in the sense of {\it plectics} \cite{GellMann94}, belong to the world within which the concepts used in $BG$ statistical mechanics have, since more than one century,  been profusely shown to be the appropriate ones. 

The cases (iii) and (iv), which we may call {\it complex} in the sense of {\it plectics} \cite{GellMann94}, belong to the world within which the concepts used in nonextensive statistical mechanics (as well as in its possible generalizations or alternatives) have been shown and keep being shown (since more than one decade, by now) to be the appropriate ones.

The cause for a system to be {\it simple} or {\it complex} in the above sense lies basically (see \cite{Einstein10, Cohen04}) on its microscopic dynamics in the {\it full} space of microscopic possibilities (Gibbs' $\Gamma$-space for many-body Hamiltonian systems). It is believed to be so because it is this dynamics which is expected to determine the possible persistent correlations that would be responsible for the geometrical structure within which the system tends to ``live", given its initial conditions. 

If its elementary nonlinear dynamics is controlled by {\it strong} sensitivity to the initial conditions (i.e., at least one {\it positive} Lyapunov exponent if the system is a classical one), we expect the system to be {\it simple}. For virtually any initial condition, the system quickly visits the neighborhood of virtually all possible states; it does so in such a way that the probability of any small part of the (macroscopically) admissible full space asymptotically becomes proportional to its size (Lebesgue measure). Its main time-dependent functions (typically) {\it exponentially} depend on time, the number of allowed stated {\it exponentially} increases with $N$, and, if the system is Hamiltonian, its stationary state (called {\it thermal equilibrium}) is characterized by an energy distribution given by the celebrated $BG$ weight. Summarizing, its paradigmatic ordinary differential equation is $dy/dx \propto y$.

If the elementary nonlinear dynamics of the system is nonregular and controlled by {\it weak} sensitivity to the initial conditions (i.e., {\it no positive} Lyapunov exponents if the system is a classical one), we expect the system to be {\it complex}. For given initial conditions, the system essentially visits a network of states (a {\it scale-free network} for many if not all  $q$-systems) whose typical Lebesgue measure is zero. The particular network depends from the initial conditions and is highly inhomogeneous (like, as mentioned before, the network of airports on which a specific air company operates), but its geometry (both topology and metrics) is basically the same for virtually all initial conditions. To recover a homogeneous occupancy of the full space we are obliged to make {\it averages over all the initial conditions} (whereas no such thing is necessary for {\it simple} systems).
The main time-dependent functions of the system typically depend on time slower than exponentially (typically like {\it power-laws}, more precisely like $q$-exponentials for most if not all the $q$-systems), the number of allowed states  increases with $N$ like a power-law, and, if the system is Hamiltonian, its stationary or quasi-stationary (metastable) state (out of thermal equilibrium) is expected to be characterized by an energy distribution given by a $q$-exponential weight, which asymptotically approaches a power-law for high energies. Summarizing, its paradigmatic ordinary differential equation is $dy/dx \propto y^q$. 

The standard central limit theorem plays a crucial role for the {\it simple} systems. We expect a similar theorem to exist for the {\it complex} systems of the $q$-class. The proof of such a theorem would be priceless.

In addition to the above, we have presented here more two  conjectures (Eqs. (92) and (94)) concerning the entropy production per unit time for a nonlinear dynamical system at the edge of chaos, having $\mu$ vanishing Lyapunov exponents. The entropy $S_q$ which increases {\it linearly} with time (when the system is exploring its phase space) appears to be that whose entropic index is $q_e$ as given  by Eq. (92) \cite{AnanosBaldovinTsallis04}. The associated entropy production $K_{q_e}$ per unit time appears to be as given by Eq. (94). The proof of these two connected conjectures (including, naturally, the precise conditions for their validity) also remains as a mathematically open problem.

\section*{Acknowledgments}

We acknowledge useful discussions with S. Abe, C. Anteneodo, F. Baldovin, S. Curilef, M. Gell-Mann, R. Hersh, L. Moyano, C. Pagani and Y. Sato, as well as partial financial support by Pronex/MCT, Faperj and CNPq (Brazilian agencies).


\begin{thebibliography}{10}
\bibitem{remark}A simple generalization also covering possibly negative real numbers $X$ and $Y$  is the following one:
$X \otimes_q Y \equiv [sign(X)][sign(Y)][|X|^{1-q}+|Y|^{1-q}-1]^{1/(1-q)}$.
\end{thebibliography}
\end{document}